\documentstyle[12pt,aasms4]{article}

\received{}
\accepted{}

\slugcomment{Submitted to the Astrophysical Journal Letters}

\lefthead{Wilson \& Raymond}
\righthead{Photoionizing Shocks in Seyfert Narrow Line Regions}

\begin{document}

\title{Do Jet-Driven Shocks ionize the Narrow Line Regions of Seyfert Galaxies?}

\author{A. S. Wilson\altaffilmark{1}}
\affil{Space Telescope Science Institute, 3700 San Martin Drive, Baltimore, MD
21218; awilson@stsci.edu}

\and

\author{J. C. Raymond}
\affil{Harvard-Smithsonian Center for Astrophysics, 60 Garden Street,
Cambridge, MA 02138; raymond@cfa.harvard.edu}


\altaffiltext{1}{Also Astronomy Department, University of Maryland,
College Park, MD 20742; wilson@astro.umd.edu}


\begin{abstract}

We consider a model in which the narrow line regions (NLRs) of Seyfert
galaxies are photoionized ``in situ'' by fast (300 -- 1,000 km s$^{-1}$),
radiative shock waves driven into the interstellar medium of the galaxy
by radio jets from the active nucleus.
Such shocks are powerful sources of soft X-rays. We
compute the expected ratio of the count rates in the ROSAT PSPC and
Einstein IPC detectors to the [OIII]$\lambda$5007 flux as a function of
shock velocity, and compare these ratios with observations of type 2 Seyferts.
{\it If} most of the observed soft X-ray emission from these galaxies originates
in the NLR and the absorbing hydrogen column is similar to that inferred from
the reddening of the NLR, a photoionizing shock model with shock velocity
$\simeq$ 400 -- 500 km s$^{-1}$ is compatible with the observed ratios. High
angular resolution observations with AXAF 
are needed to isolate the X-ray emission of the
NLR and measure its absorbing column, thus providing a more conclusive test.
We also calculate the expected coronal iron line emission from the shocks.
For most Seyfert 2s, the [Fe X]$\lambda$6374/H$\beta$ ratio is a factor of 
2 -- 14 lower than the predictions of 300 -- 500 km s$^{-1}$ shock models,
suggesting that less hot gas is present than required by these models.
\end{abstract}




%

\newpage
\section{Introduction}

The gas in the narrow line regions (NLRs) of Seyfert galaxies is known
to be photoionized (e.g. Koski 1978; Ferland \& Osterbrock 1986;
Osterbrock 1989) and the prevailing view is that the ionizing photons
originate in a compact source, perhaps the accretion disk.
However, strong associations between the narrow emission-line
and radio continuum properties of Seyfert nuclei have been found over the
last 20 years. There are strong correlations between radio power and
both line luminosity (de Bruyn \& Wilson 1978) and line width
(Wilson \& Willis 1980). Early VLA maps showed that the radio sources represent
collimated ejection from the compact nucleus and that they have similar
spatial scales and orientations to the NLR. These results led Wilson \& Willis
(1980) to
suggest that the nucleus ejects radio components which interact with ambient
gas and replenish the high kinetic energy and ionization of the NLR.
Simple models
of the momentum transfer between jets and ambient gas (Blandford \&
K\"onigl 1979)
show that ambient gas with the observationally-inferred 
NLR mass can be accelerated to the
observed velocities ($\sim$ 10$^{3}$ km s$^{-1}$) in the gas crossing
time of the NLR ($\sim$ 10$^{5-6}$ yrs) for reasonable efficiency factors
relating jet and radio powers (Wilson 1981, 1982). Subsequent imaging
(e.g. Haniff et al. 1988; Bower et al. 1995;
Capetti et al. 1995; Falcke, Wilson \& Simpson 1998) and
spectroscopic (Whittle et al. 1988; Axon et al. 1998)
observations have confirmed that the structure of the NLR in many Seyferts
is dominated by compression of interstellar gas by 
the radio ejecta and that these ejecta are an important source
of ``stirring'' of the gas. Theoretical descriptions of this interaction have
involved expanding radio lobes (Pedlar, Dyson \& Unger 1985), bow shocks
driven by the radio jets (Taylor, Dyson \& Axon 1992; Ferruit et al. 1997)
and the role of the jet cocoon (Steffen et al. 1997).

The power required to accelerate the gas is typically 
$\sim$ 10$^{41-42}$ erg s$^{-1}$
over $\sim$ 10$^{5-6}$ yrs to give the $\sim$ 10$^{53-55}$ erg of kinetic
energy present in the observed clouds of the NLR (more kinetic energy could
be present in gas too tenuous to be visible in line emission).
The power radiated in observed line emission is somewhat
higher at $\sim$ 10$^{41-44}$
erg s$^{-1}$, which is $\sim$ 10$^{2-4}$ times higher than the radio
luminosity (e.g. Fig. 1 of Wilson, Ward \& Haniff 1988).
Recently, Bicknell et al. (1998) have argued that the narrow
line emission in Seyfert galaxies is, indeed, powered entirely by the
radio-emitting jets. 
The jets are
considered to drive several hundred to a thousand km s$^{-1}$ radiative 
shocks into interstellar gas on the hundred pc scale.
Such ``photoionizing shocks'' are
powerful sources of ionizing radiation and create photoionized precursors
(Daltabuit \& Cox 1972),
the optical spectra of which are similar to Seyfert galaxy NLRs
(Dopita \& Sutherland 1995, 1996 [hereafter DSI];
Morse, Raymond \& Wilson 1996). In this
picture, the conversion of jet kinetic energy to radio emission is much
less efficient in Seyfert galaxies than in radio galaxies and radio-loud
quasars; the Seyfert jets are thermally dominated while jets in radio-loud
AGN may be dominated by relativistic particles and magnetic fields.

It is very important to decide whether the energy source that powers the
NLR is ``in situ'' mechanical motion (jets) or the more conventional compact
photoionizing source. This issue is closely related to the fundamental
question of whether the putative
accretion disk loses energy primarily in mechanical (as jets or winds)
or radiative form.

Shock velocities of 300 - 1,000 km s$^{-1}$, required to account for
the emission-line spectra of Seyferts (Dopita \& Sutherland 1995), 
generate gas of temperature 10$^{6-7}$K and copious soft
X-rays (cf. Laor 1998). This production of soft X-rays by the NLR
is an inevitable prediction of the photoionizing shock model
and can be used to distinguish it from 
photoionization by a hidden Seyfert 1
nucleus. Indeed, soft X-ray emission from the NLR has been isolated in
NGC 1068 (Wilson et al. 1992), NGC 2110 (Weaver et al. 1995) and NGC 4151
(Morse et al. 1995). For type 2 Seyfert galaxies, in general,
the observed Einstein IPC soft X-ray flux (from the {\it entire} galaxy)
and the [OIII]$\lambda$5007
flux (from the NLR) are similar (Fig. 1).
This similarity is not inconsistent with
photoionizing shocks being the power source, since $\simeq$ 4\% (20\%)
of the kinetic power of a 300 (500) km s$^{-1}$ radiative shock is radiated in
the Einstein band (0.2 -- 4 keV), while $\simeq$ 2\% is radiated as
[OIII]$\lambda$5007
(DSI; Bicknell, Dopita \& O'Dea 1997, hereafter BDO), and
the expected X-ray spectrum is soft and
photoelectric absorption can substantially reduce the observed flux.
In this letter, we examine the relationship between soft X-ray and
[OIII]$\lambda$5007 emission expected in the photoionizing shock model, and
compare it with Einstein and ROSAT X-ray observations.
We also argue that the shocks should produce strong coronal iron line
emission, and compare the predicted and observed strengths.

\section{Comparison of [OIII]$\lambda$5007 and Soft X-ray Fluxes}

\subsection{Method}

The calculations of DSI provide the luminosity radiated
in the [OIII]$\lambda$5007 line per unit area of shock 
from both the post-shock gas and the
photoionized precursor as a function of shock velocity and magnetic parameter.
The pre-shock density is taken to be low enough that collisional
deexcitation is unimportant. 
We have used the model-predicted [OIII]$\lambda$5007 luminosities from DSI
(with the fluxes reduced by a factor of 2,
as we suspect they may be too high by this factor based on comparison with
the models discussed below and an apparent factor of 2 error in 
equation 3.3 of DSI) and BDO.
For the post-shock gas, the [OIII]$\lambda$5007 luminosity
was taken to be the average of the predictions for the four magnetic
parameters considered by DSI. 
The exact value of the magnetic parameter is not
critical because for shock velocities above $\simeq$ 200 km s$^{-1}$, the
[OIII] emission of the precursor dominates that of the post-shock gas.
The [OIII]$\lambda$5007 luminosity would be less than predicted if any of the
following effects are important: a)
the precursor is density bounded or the pre-shock gas presents a covering
factor of $<$ 1; b) the post-shock gas cools in an unstable fashion, so
that the fragmented gas does not intercept all of the ionizing flux
(e.g. Innes, Giddings \& Falle 1987a, b); or c) the pre-shock density is
sufficiently high that the $^{1}D_{2}$ level of OIII suffers collisional
deexcitation. Of these, effect b) is unlikely to affect our results 
significantly in 
view of the small contribution of the post-shock gas to
[OIII]$\lambda$5007 for the shock velocities of interest. Since the
collisional deexcitation density of the $^{1}D_{2}$ level is 7.0 $\times$
10$^{5}$ cm$^{-3}$ (Osterbrock 1989),
it is very unlikely that effect c) applies to the precursor gas,
though post-shock gas could be collisionally deexcited
in the inner, denser parts of the NLR. The possibility that effect a) is
significant means that the predicted [OIII]$\lambda$5007 luminosities
are most conservatively treated as upper limits.

We have also calculated the luminosity of the 
radiation emitted from the post-shock
gas in the bands of the ROSAT PSPC (0.1 - 2.4 keV) and Einstein
IPC (0.2 - 4 keV) detectors using an updated version of the radiative
shock wave code described in Raymond (1979).  The atomic rates are
basically the same as those of the current version of the Raymond \&
Smith (1977) X-ray code used for the X-ray spectrum predictions in
DSI.  The shock code includes time-dependent ionization and photoionization,
but these have little effect on broad-band X-ray count rates.
To obtain the total emission (over 4$\pi$ sterad), we multiplied the X-ray
flux radiated upstream by a factor of 2. This factor may be too large for
soft photons because of photoelectric absorption by the downstream gas.
The predicted ratio, R,
of X-ray count rate to [OIII]$\lambda$5007 flux can then be obtained as a
function of shock velocity and effective hydrogen column density.
The
code also predicts the intensities of the [Fe X] and [Fe XIV] lines
discussed below, including excitation by electron and proton impact,
and by cascades following LS permitted excitations.

\subsection{Results}

Figs 1 and 2 show the Einstein IPC and ROSAT PSPC count rates against the
[OIII]$\lambda$5007 flux.
The dashed
lines in each panel are loci of constant R obtained from the model
as described above after attenuation
by an equivalent hydrogen column of
N$_{H}$ = 10$^{20.5}$ cm$^{-2}$, which is the average Galactic column density
towards the Seyfert 2 galaxies plotted in the figures (see below).
The [OIII] fluxes have been reduced by 0.2 mag, which is the expected 
obscuration for a normal gas to dust ratio.
The solid lines are the same but with the model X-ray fluxes
attenuated by N$_{H}$ = 10$^{21.5}$ cm$^{-2}$, which is
the column density corresponding to the average obscuration
(A$_{V}$ = 1.8 mag, derived from the
Balmer decrement assuming intrinsic case B recombination values) of
the NLRs, assuming a normal gas to dust ratio.
The (narrow)
hydrogen lines in Seyfert 2s come from a similar region to [OIII]$\lambda$5007,
so the model-predicted 
[OIII]$\lambda$5007 fluxes have been reduced by 1.12 $\times$ A$_{V}$ =
2.0 mag.
Because the model [OIII]$\lambda$5007 fluxes
may be overestimates, as discussed above, the predicted values of R are best
treated as {\it lower limits}, so the
predicted locations in Figs 1 and 2 for a given shock velocity lie to the
upper left of the plotted line.

Also plotted in Figs 1 and 2 are observed X-ray count rates and 
[OIII]$\lambda$5007 fluxes for type 2 Seyfert galaxies. Type 1 Seyferts
and the so-called ``narrow line X-ray galaxies'' have been omitted (with
the exception of NGC 2110, see figure captions) because the equivalent
hydrogen columns to their compact nuclei
are sufficiently low that the soft X-ray
emission is dominated by the Seyfert 1 nucleus. The
equivalent hydrogen column towards the ``hidden'' Seyfert 1 nuclei in
Seyfert 2 galaxies averages $\simeq$ 3 $\times$ 10$^{23}$ cm$^{-2}$
(Turner et al.
1997), so the compact source is heavily attenuated in the Einstein and
ROSAT bands. At the spatial resolution of the IPC and PSPC detectors, 
the measured X-ray count rate generally includes the entire host galaxy,
the NLR and any residual transmitted or scattered emission from the compact
nucleus. Thus these
observed X-ray fluxes represent firm upper limits to the X-ray
flux from the NLR. The [OIII]$\lambda$5007 emission, on the other hand,
was measured through a small aperture and is a good measure of the
NLR line emission. Thus the plotted data represent {\it upper limits} to
the actual value of R in the NLR.

Also plotted in Fig. 2 is a point representing 
the total PSPC count rate (after absorption)
and [OIII]$\lambda$5007 flux density (after obscuration)
for a 3$^{\prime\prime}$ (0.7 pc) length along the shock front
of the LMC supernova remnant
N132D (Morse et al. 1996). The plotted PSPC X-ray count rate was
obtained by integrating
the curve of 0.1 - 2 keV flux density (obtained from the ROSAT HRI image
and corrected for absorption) versus distance 
perpendicular to the shock given in Fig. 11
of Morse et al. (1996).
The predicted absorbed PSPC flux density was then found using the spectral
model of Hwang et al. (1993) and the equivalent hydrogen column of 
N$_{H}$ = 6.2 $\times$ 10$^{20}$ cm$^{-2}$, which is within a factor of 2 of the
column used for the models represented by the dashed lines. The curve of 
[OIII]$\lambda$5007 flux density versus distance perpendicular to the shock,
also given in Fig. 11 of Morse et al. (1996), was similarly integrated and was
reduced by an obscuration A$_{V}$ = 0.34 mag
(equivalent to N$_{H}$ = 6.2 $\times$ 10$^{20}$ cm$^{-2}$ for a normal gas to
dust ratio). Fig. 2 shows that the resulting value of R for N132D
is about 50 times higher than the average of the Seyfert galaxy points.
Morse et al. (1996) argue convincingly
that the X-ray and [OIII]$\lambda$5007 emissions from this region of N132D
originate from a 
photoionizing shock of velocity 790 km s$^{-1}$, with some contribution to the
ionizing flux from lower velocity shocks. The absorption-corrected ratio
F(0.1 - 2 keV)/F([OIII]$\lambda$5007) for 
this 0.7 pc length of shock is $\simeq$ 45, with an uncertainty of about 25\%.

Examination of Figs 1 and 2 shows that the model generally
overpredicts the observed
R values (i.e. the predicted X-ray flux is too high) for high shock velocities
(500 - 1,000 km s$^{-1}$)
when only the Galactic column (N$_{H}$ =
10$^{20.5}$ cm$^{-2}$) is included. 
For this
column density, a shock velocity of $\sim$ 300 km s$^{-1}$ better describes
the observed R values.

For N$_{H}$ = 10$^{21.5}$ cm$^{-2}$, the model of a 500 km s$^{-1}$ shock
is comparable to
most observed detections and upper limits, while the X-ray
emission of a 300 km s$^{-1}$ shock is $\sim$ 2 orders of magnitude (the exact
number being very sensitive to N$_{H}$) below the observed data points.
{\it If
most of the observed soft X-ray emission originates in the NLR} and the hydrogen
column is similar to that inferred towards the optical (narrow) line emission,
a photoionizing shock model with shock velocity
$\simeq$ 400 -- 500 km s$^{-1}$ is compatible with the observations. On the
other hand, if the X-ray
flux from the NLR is actually substantially lower than the observed total
emission from the galaxy, a lower value of the shock velocity would be
indicated or a classical photoionization model favored.
Future high spatial resolution, medium spectral resolution 
observations with the ACIS on AXAF will spatially resolve the
X-ray emission of the NLR in nearby Seyferts, and thus separate it from 
both the galaxy disk and the compact nucleus.
These measurements will also
allow the gas temperature and N$_{H}$ to be determined, providing a
much more precise check of the model.

\section{Coronal Iron Line Emission}

As emphasised by DSI, the post-shock gas should display a rich collisionally
ionized UV spectrum. However, these lines are very sensitive to reddening. 
Another prediction of fast shock models is strong
emission in the coronal forbidden lines of iron, such as [Fe X]$\lambda$6374 
and [Fe XIV]$\lambda$5303. These lines are expected whenever the shocked gas
is hot enough
to collisionally ionize Fe to these species. 
Table 1 shows both absolute fluxes and fluxes relative to
H$\beta$ in these two lines. The H$\beta$ emission (H$\beta_{S}$) is that
from only the
post-shock gas - emission from the precursor is not included in the listed
ratio. The range of
shock velocities consistent with both the optical line ratios observed in
Seyferts
(Dopita \& Sutherland 1995) and the soft X-ray observations (Section 2)
is 300 - 500 km s$^{-1}$. For such shock velocities, the ratio of
total H$\beta$ flux (H$\beta_{T}$ - shock plus precursor) to shock-only flux
(H$\beta_{S}$)
is 2.1 - 2.2 (DSI, tables 8 and 10). Thus the predicted 
[Fe X]$\lambda$6374/H$\beta_{T}$ ratios are in the range 0.1 - 0.19, excluding
any [Fe X]$\lambda$6374 from the photoionized precursor. The observed,
reddening-corrected
[Fe X]$\lambda$6374/H$\beta$ ratios in Seyfert 2s are in the range
$<$ 0.014 - 0.16 (Koski 1978; Penston et al. 1984), with most galaxies
in the range $<$ 0.014 - 0.06. The strongest [Fe X]$\lambda$6374
emitters are consistent with a 300 - 500 km s$^{-1}$ shock model,
but most galaxies
show an [Fe X]$\lambda$6374/H$\beta$ ratio which is a factor of 2 - 14 lower
than the predictions. A similar conclusion comes from
the general lack of detection of [Fe XIV]$\lambda$5303 in Seyfert
galaxies. Of course, a contribution from lower velocity ($<$ 300 km s$^{-1}$)
shocks would enhance the H$\beta$ flux without significantly increasing 
[Fe X]$\lambda$6374 or [Fe XIV]$\lambda$5303 and thus reduce the
predicted ratio. However, the optical emission line spectrum would
then be of low excitation and would not match a Seyfert NLR spectrum.
Thus the weakness of the iron coronal lines may favor a conventional
photoionization model over the shock model.
Depletion of iron onto
grains would reduce the coronal line fluxes, but it would decrease the
X-ray emissivity as well, and much of the iron is liberated by the time
radiative cooling sets in (Vancura et al. 1994).  Thus the [Fe X]/H$\beta$
ratios would be reduced by a modest amount in models with grains.

This research was supported by NASA through grants NAG 53393 and NAG 81027, and
by NSF through grant AST9527289. We are grateful to P. Ferruit, J. A. Morse
and N. Nagar for help and advice.

\vfil\eject

\clearpage

\figcaption[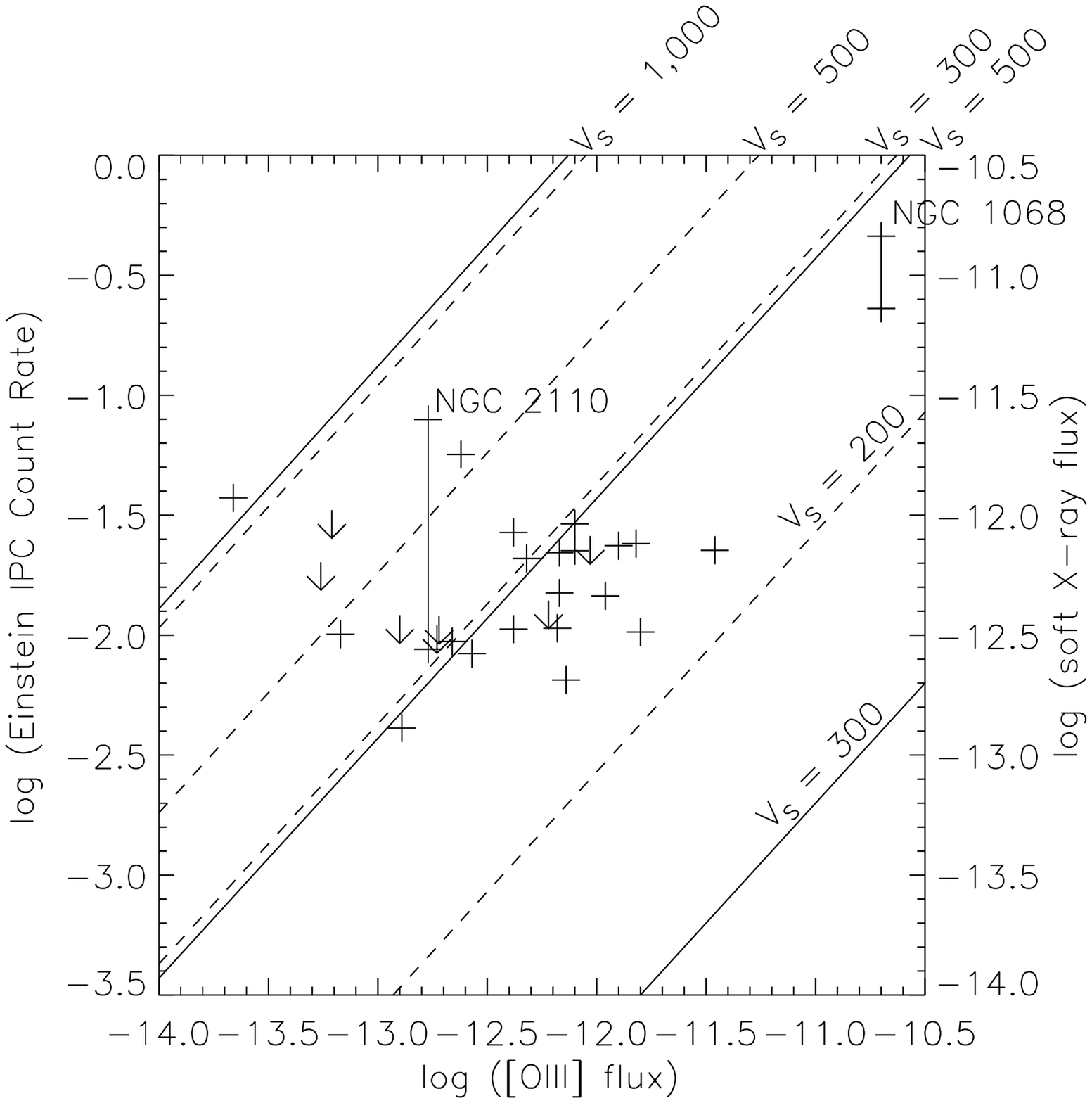]
{A plot of the logarithm of the total Einstein IPC
count rate (in counts s$^{-1}$, left axis)
versus the logarithm of the flux
in [OIII]$\lambda$5007 (in erg cm$^{-2}$ s$^{-1}$). 
The right axis gives the corresponding logarithm of the soft X-ray flux
(in erg cm$^{-2}$ s$^{-1}$ in the 0.2 - 4.0 keV band, with no correction for
absorption). The points and upper limits are data for Seyfert 2 galaxies from 
Mulchaey et al. (1994). With the exception of NGC 2110, so called
``narrow line X-ray galaxies'' are omitted.
The vertical line descending
from the upper NGC 2110
point joins the total observed count rate to the approximate count rate of
extended X-ray emission some 4$^{\prime\prime}$ N of the nucleus and
within the NLR
(from Weaver et al. 1995). The vertical line descending from the upper
NGC 1068 point joins the observed count rate from the entire galaxy
to the approximate count
rate of the spatially resolved emission in the ROSAT HRI image
(from Wilson et al. 1992). The diagonal straight lines 
are predictions of photoionizing shock models, using calculations of
[OIII]$\lambda$5007 flux from either
DSI (modified as mentioned in the text and used for shock velocities
$\le$ 500 km s$^{-1}$) or BDO (for a shock velocity of 1,000 km s$^{-1}$), 
and calculations of
the Einstein IPC count rate described in the text. The shock
velocities (in km s$^{-1}$) are indicated.
The dashed lines show the model predicted Einstein
count rates reduced by the effects of photoelectric absorption by
an equivalent hydrogen column of N$_{H}$ = 10$^{20.5}$ cm$^{-2}$
(which is the mean
Galactic column to the galaxies plotted), and the model predicted
[OIII]$\lambda$5007 fluxes reduced by an obscuration
A$_{5007}$ = 0.2 mag. The solid lines are the
predictions of the model for an absorbing column
of N$_{H}$ = 10$^{21.5}$ cm$^{-2}$ (which is the mean
column inferred from the Balmer decrement of the NLR) and an
obscuration A$_{5007}$ = 2.0 mag.
\label{Figure 2}}

\figcaption[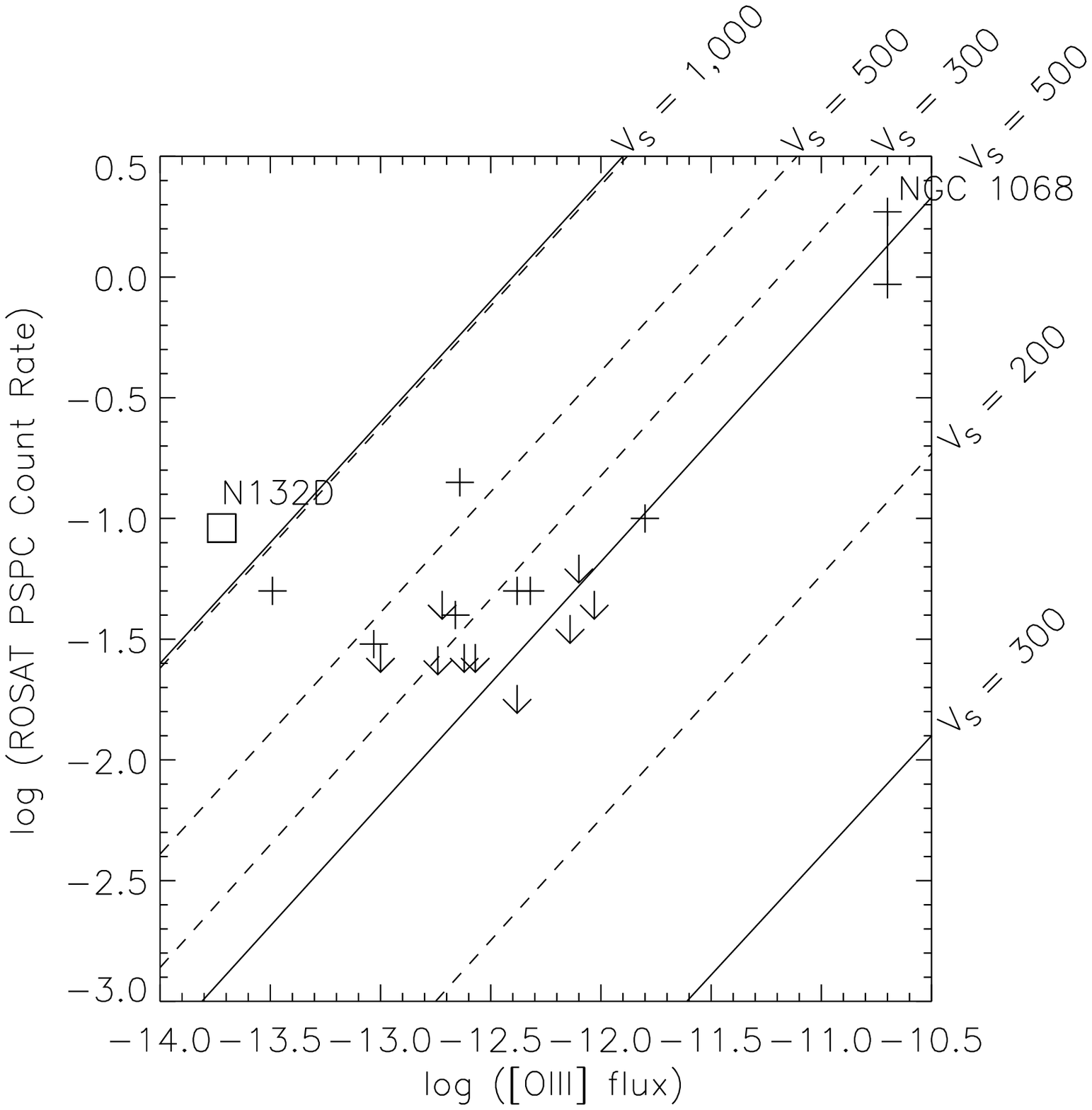]
{As Fig. 1, but for the ROSAT PSPC count rate. The data plotted are the
Seyfert 2 galaxies in Rush et al. (1996), with [OIII]$\lambda$5007 fluxes
from Whittle (1992). The rectangle represents a 0.7 pc
length along the shock front of the LMC SNR N132D (see text).
\label{Figure 3}}

\clearpage

\begin{deluxetable}{lcccc}
\tablecolumns{5}
\tablewidth{0pc}
\tablecaption{Coronal Line Emission from the Shock}
\tablehead{
Shock velocity (km s$^{-1}$) & \colhead{[Fe X]$\lambda$6374}
& \colhead{[Fe XIV]$\lambda$5303} &\colhead{[Fe X]$\lambda$6374/H$\beta_{S}$}
& \colhead{[Fe XIV]$\lambda$5303/H$\beta_{S}$}\\ }
\startdata
200 & 2.16$\times$10$^{-8}$ & 0.0                    & 0.0025 & 0.0   \nl
300 & 8.10$\times$10$^{-6}$ & 1.90$\times$10$^{-7}$ & 0.41   & 0.01   \nl
350 & 1.19$\times$10$^{-5}$ & 9.38$\times$10$^{-6}$ & 0.38   & 0.30   \nl
400 & 1.41$\times$10$^{-5}$ & 3.85$\times$10$^{-5}$ & 0.35   & 0.98   \nl
500 & 1.64$\times$10$^{-5}$ & 7.87$\times$10$^{-5}$ & 0.23   & 1.09   \nl
1,000 & 3.34$\times$10$^{-5}$  & 1.52$\times$10$^{-4}$ & 0.13  & 0.62  \nl
\enddata
\label{tab:comp-pos}
\tablenotetext{} {The numbers in columns 2 and 3 represent the absolute fluxes
(in erg cm$^{-2}$ s$^{-1}$) in the coronal lines emitted per unit area of shock
for a pre-shock density of 1 atom cm$^{-3}$.}
\end{deluxetable}

\end{document}